\documentclass[prb,amsmath,amssymb,floatfix]{revtex4}
\usepackage{graphicx}

\newcommand{\be}{\begin{eqnarray}}
\newcommand{\ee}{\end{eqnarray}}
\newcommand{\bfr}{{\bf r}}
\newcommand{\bfq}{{\bf q}}
\newcommand{\bfp}{{\bf p}}
\newcommand{\bfk}{{\bf k}}

\newcommand{\bfR}{{\bf R}}

\newcommand{\bfP}{{\bf P}}

\newcommand{\hata}{\hat{a}}
\newcommand{\hatb}{\hat{b}}
\newcommand{\hatr}{\hat{r}}
\newcommand{\hatk}{\hat{k}}
\newcommand{\hatp}{\hat{p}}

\newcommand{\hatpsi}{\hat{\psi}}
\newcommand{\wbe}{\begin{widetext}}
\newcommand{\wee}{\end{widetext}}
\newcommand{\oncite}{\onlinecite}

\begin{document}
\draft

\title{An effective many-body theory for strongly interacting polar molecules}

\author{Daw-Wei Wang}

\address{Physics Department, and
National Center for Theoretical Science, 
National Tsing-Hua University, Hsinchu, Taiwan, ROC}

\date{\today}

\begin{abstract}
We derive a general effective many-body theory for 
bosonic polar molecules in strong interaction regime, which 
cannot be correctly described by previous theories within 
the first Born approximation.  
The effective Hamiltonian has additional interaction terms, which
surprisingly reduces the anisotropic features of dipolar interaction
near the shape resonance regime.
In the 2D system with dipole moment perpendicular to the plane, 
we find that the phonon dispersion scales
as $\sqrt{|\bfp|}$ in the low momentum ($\bfp$) limit,
showing the same low energy properties as a 2D charged Bose gas
with Coulomb ($1/r$) interactions.
\end{abstract}

\maketitle

\section{Introduction}

Recent developments in the trapping and cooling of chromium atoms
[\onlinecite{Cr}] and polar molecules [\onlinecite{cool_molecule}]
provide a new direction for investigating quantum states
resulting from the anisotropic dipole
interaction. Dipolar effects on the 
condensate profile [\oncite{dipole_size}] and 
elementary excitations [\oncite{dipole_excitation}]
have been extensively studied both theoretically and 
experimentally. Several exotic many-body states resulted from 
dipolar interactions are also proposed 
[\oncite{dipole_exotic}]. However, most 
theoretical works so far are based on the pseudo-potential
developed by Yi and You [\oncite{You_pseudo}] 
within the first Born approximation (FBA).
As a result, these results become not justified when applied to
the systems of polar molecules, which can have large electric
dipole moments and hence strong dipolar interaction to renormalize
the scattering amplitude beyond the FBA.

In Ref. [\oncite{Andrey}], Derevianko extended Huang and Yang's 
approach [\oncite{Yang}] to the anisotropic dipolar interaction,
and for the first time shew how it may be possible to go beyond 
the Born approximation in a dipolar gas system. The derived pseudo-potential,
however, is non-hermitian in the low energy limit,
and therefore cannot be easily used for constructing the effective theory 
of strongly interacting dipolar gases.
In Ref. [\oncite{Bohn_prl}], the authors studied the systems of 
bosonic dipoles via Monte Carlo calculation, and found that the 
ground state energy can be well-explained 
by the Yi and You's pseudo-potential within FBA if using  
a dipole-dependent $s$-wave scattering. Results of Ref. [\oncite{Bohn_prl}]
can certainly be applied to a regime of stronger dipole momentum
(beyond the valid regime of Yi and You's original pseudo-potential), 
because the higher order renormalization of the $s$-wave scattering amplitude
has been included. But its validity to apply to polar molecules with
large dipole moment in strong field is still questionable,
because the higher order renormalization to the scattering amplitude
(FBA is the first order perturbation) of the {\it non-$s$-wave} scattering
channels are not included at all. As a example, in Ref. [\oncite{You_Born1}],
Deb and You found that the scattering matrix element between $s$-wave
and $d$-wave channels also has strong deviation from their weak interaction
result when near the shape resonance. It is reasonable to believe that
there will be such kind of deviation from the FBA results in other
channels in stronger dipole moment regime, as usually considered in the
polar molecule systems.
Therefore developing a correct and widely-applicable pseudo-potential 
and the associated many-body theory for systems of
dipolar atoms/molecules is still a very important and crucial 
step for future theoretical and experimental studies.

In this paper, we derive a complete effective many-body theory which
can correctly describe bosonic polar molecules both
in the weak and strong interaction regime and/or near 
the SR. The resulting effective Hamiltonian 
is modified by additional three-point and four-point interactions,
which may significantly change the nature of condensate profile/dynamics.
For example, when the dipole moment is 
near the first $s$-wave shape resonance regime,
we find that the additional interaction (new terms beyond the FBA) 
can reduce the anisotropy of the condensate profile. 
In a 2D uniform system with the dipole moment perpendicular
to the plane, we find that the phonon dispersion
scales as $\sqrt{|\bfq|}$ (instead of $|\bfq|$ in the typical Bogoliubov
mode) in the long wavelength limit, showing the same low energy physics as
2D charged bosons with Coulomb ($1/r$) interactions
[\oncite{2D_C_BEC}]. As a result, our theory is important not only
in the study of strongly interacting polar molecules, 
but also in the possible application of simulating the liquid phase
of 2D charged bosons by neutral particles. Such simulation cannot be 
done in ion traps because of the strong Coulomb potential compared to 
the kinetic energy. These results may be useful in studying the
properties of High $T_c$ superconducting thin film, where
the coherent length of Cooper pairs are known to be very small as 
a composite charged boson [\oncite{high_Tc}].

The paper is organized as following: In Sec. \ref{scattering},
we first discuss the general 
scattering theory of dipolar interaction and briefly review theories used in
previous work. In Sec. \ref{FBA} we used a exactly
solvable model to exam the validity of the first
Born approximation of dipole interaction. In Sec. \ref{3D}, we derive
the correct effective theory and the associated Gross-Pitaivskii equation
for bosonic polar molecules in 3D system. We then 
discuss the condensate wavefunction
by using variational method in Sec. \ref{variational}. Finally, we extend
the 3D results to develop an effective theory in quasi-2D system in 
Sec. \ref{2D}, and calculate the phonon mode dispersion as well as the 
Kosterlitz-Thouless transition temperature. We then summarize 
our work in Sec. \ref{summary}.

\section{Low energy scattering theory of dipoles}
\label{scattering}

For the convenience of later discussions, we first briefly
review the recent progress on the scattering problem of dipolar gas,
where the electric/magnetic dipole moment is polarized by the external
electric/magnetic field along $z$ direction.
The most general form of the scattering amplitude between two identical
particles in such situation can be expressed to be 
\be
f(\bfk,\bfk')=4\pi\sum_{lm}\sum_{l'm'}t_{lm}^{l'm'}(k)Y_{lm}^\ast(\hat{k})
Y_{l'm'}^{}(\hat{k}'),
\label{f_general}
\ee
where the scattering matrix element, $t_{lm}^{l'm'}(k)$, depends
on the relative incident momentum, $k$, and the summation is over even
$l$ for bosons and odd $l$ for fermions. $Y_{lm}(\hatr)$ is
the spherical harmonic function of unit vector $\hatr$.
At large distances, the inter-particle
potential is dominated by the dipolar term, 
$V_{\rm d}(\bfr)=D^2(1-3\cos^2\theta)/|\bfr|^{3}$, where 
$D$ is the electric dipole moment in c.g.s. unit (for simplicity,
here we use electric dipoles to formulate the theory for polar molecules,
while an similar version for magnetic dipolar gas can be also obtained
easily); $\theta$ is the angle between the distance $\bfr$ and 
the dipole direction (polarized in $z$ direction).
However, at short distances the potential becomes much more complicated
due to the Coulomb and spin exchange interaction between electrons.
Deb and You [\oncite{You_Born1}] first calculated the cross section
within a certain model potential and studied how they are changed near
the shape resonance regime. Based on the numerical results,
Yi and You [\oncite{You_pseudo}] then proposed a pseudo-potential: 
\be
V_{\rm ps}(\bfr)\equiv \frac{4\pi a_{s}\hbar^2}{M}\delta(\bfr)
+V_{\rm d}(\bfr),
\label{V_ps}
\ee
to calculate the {\it low energy} scattering matrix element 
within the first Born approximation {\it away} from the shape resonance 
regime. Here $a_s=-t_{00}^{00}(0)$ is the $s$-wave scattering length 
in zero field limit.
Within the FBA [\oncite{dipole_size,dipole_excitation}],
$f_{B}(\bfk,\bfk')=\frac{-M}{4\pi\hbar^2}\int d\bfr\,
e^{i(\bfk-\bfk')\cdot\bfr}V_{\rm ps}(\bfr)
=-a_s-\frac{2}{3}a_dP_2\left(\cos\theta_{\bfk-\bfk'}\right)$,
where $a_d\equiv MD^2/\hbar^2$ is a length scale
and $P_2(x)$ is the Legendre Polynomial. $\theta_\bfk$ is the angle between
the momentum $\bfk$ and $z$ axis.
As a result, the associated matrix elements become:
${t_B}{}_{lm}^{l'm'}(k)=4\sqrt{\frac{\pi}{5}}a_d i^{l-l'}$
$\int d\Omega_r Y_{lm}(\hatr)Y^\ast_{l'm'}(\hatr)Y_{20}(\hatr)
\int_{kr_c}^\infty \frac{dr}{r}j_l(r)j_{l'}(r)$ with $j_l(r)$ being
the spherical Bessel function. Here $r_c$ is a cut-off in 
the atomic length scale and therefore we can always take
$kr_c\to 0$ in the low energy limit. In above FBA result,
all the short-ranged effects are included
in the $s$-wave part ($a_s$) only, while all other matrix element, 
$t_B{}_{lm}^{l'm'}(0)$, are proportional
to the same length scale, $a_d\propto D^2$.
Therefore, it is easy to see why such results cannot be valid 
when the dipole moment (or external field) is sufficiently 
strong as higher order renormalization becomes important. 

In Ref. [\oncite{Andrey}], Derevianko developed a different
pseudo-potential for dipolar interaction to go beyond the FBA. 
Although the most general expression of the pseudo-potential is derived 
for each scattering channels and the results are in principle
applicable to strong interaction regime, but only one
terms (the scattering between $s$-wave and $d$-wave, i.e. the 
$t_{00}^{20}(0)=t_{20}^{00}(0)$ term in Eq. (\ref{f_general})) 
is evaluated within
the leading order perturbative method (equivalent to the FBA level).
In fact, we observe that Derevianko's result for the on-shell
scattering channel ($|\bfk|=|\bfk'|$) is equivalent to
the first two terms of the FBA result (i.e. the full 
$f_B(\bfk,\bfk')$ is replaced by $-a_{s}
-\frac{a_d}{6}\left[P_2(\cos\theta_{\bfk})
+P_2(\cos\theta_{\bfk'})\right]$, using
$Y_{20}(\hat{k})=\sqrt{5/4\pi}P_2(\cos\theta_\bfk)$).
This explains why
the meanfield calculation by Yi and You (Ref. [\oncite{You_PRL}], 
which included only the $s$-wave and $s$-$d$ scattering channel of
the pseudo-potential of Ref. [\oncite{Andrey}])
is not consistent with the previous result even in the weak dipole 
moment regime, where the FBA is supposed to be valid. 
(We note that this inconsistence still exist even if Yi and You
have ever used the corrected coefficient derived by the Erratum
of Ref. [\oncite{Andrey}]. The key point is that 
contributions from all other scattering
channels are all proportional to $D^2$ within the FBA 
and hence cannot be neglected compared to $t_{00}^{20}(0)$.)
In Ref. [\oncite{Bohn_prl}], Bortolotti et al. claimed that 
$V_{\rm ps}(\bfr)$ in Eq. (\ref{V_ps}) can be a good 
pseudo-potential if only
one uses a dipole-dependent $s$-wave scattering length
(i.e. $a_s(D)$). However, their
results cannot apply to the strong dipole moment regime
when the shape resonance occur in other (different from $s$-wave) 
scattering channels due to the complicated electronic density distribution 
and/or spin exchange effect in a realistic polar molecule.
Therefore, a general and useful approach to study the low energy
many-body physics of strongly interacting polar molecules is still 
needed.

\section{Criterion for the first Born approximation}
\label{FBA}

For completeness, now we explicitly examine the criterion 
for justifying the FBA in the low energy limit. 
We consider the following model potential: 
$V_{\rm mdl}(\bfr)=V_d(\bfr)$
for $|\bfr|>r_c$, and $V_{\rm mdl}(\bfr)=\infty$ for $|\bfr|\leq r_c$.
Although this model potential is over-simplified compared to the
realistic interaction potential between polar molecules, it still catches
the most important feature, anisotropic dipolar interaction, and 
hence should be useful in studying the validity of the Born approximation
in the low energy limit.
The full scattering wavefunction $\psi(\bfr)$, can be solved by:
\be
\psi(\bfr)&=&\psi_0(\bfr)-\frac{M}{\hbar^2}\int'\frac{d\bfr'}{4\pi}
G(\bfr,\bfr')V_{\rm mdl}(\bfr')\psi(\bfr'),
\ee
where $\int' d\bfr'$ is for $|\bfr'|>r_c$ only, and
\be
\psi_0(\bfr)&=&4\pi\sum_{lm}i^le^{i\delta_l(k)}\cos\delta_l(k) jn_l(k,r)
Y^\ast_{lm}(\hat{\bfk})Y_{lm}(\hat{\bfr})
\ee
is the {\it exact} scattered wavefunction 
for the hard core potential of radius $r_c$ {\it without} dipole moment.
Here we have defined $jn_l(k,r)=j_l(kr)-\tan\delta_l(k)n_l(kr)$
with $\delta_l(k)\equiv \tan^{-1}\left(j_l(kr_c)/n_l(kr_c)\right)$
being the scattering phase shift.
$j_l(x)$ and $n_l(x)$ are the conventional spherical Bessel functions.
The Green's function, $G(\bfr,\bfr')$, satisfies
$\nabla^2G(\bfr,\bfr')+k^2G(\bfr,\bfr')=-4\pi\delta(\bfr-\bfr')$ with
the boundary condition $G(r_c\hat{r},\bfr')=0$, and
therefore can be evaluated by using separation of variables.
After some straightforward calculation, the Green's function can
be expressed to be 
\be
G(\bfr,\bfr')&=&\sum_{lm}
Y^\ast_{lm}(\theta',\phi')Y_{lm}(\theta,\phi)
\frac{-4\pi k}{i+\tan\delta_l}jn_l(k,r_<)h^{(1)}_l(kr_>),
\ee
where $r_{>(<)}$ is the larger(smaller) one of $r$ and $r'$, 
and $h^{(1)}_l(x)\equiv j_l(x)+in_l(x)$. 

Within the FBA, the scattered wavefunction is 
given by the first order iteration: $\psi_B(\bfr)=
\psi_0(\bfr)-\frac{M}{\hbar^2}\int'\frac{d\bfr'}{4\pi}
G(\bfr,\bfr')V_{\rm mdl}(\bfr')\psi_0(\bfr')$. 
Therefore its validity relies on the assumption that the 
change of the wavefunction is much smaller than $\psi_0(\bfr)$ 
in the {\it whole} range of space [\oncite{Landau}].
We can therefore define a parameter, $\xi$, to measure 
the deviation of $\psi_B(\bfr)$:
$\xi\equiv\lim_{k\to 0}\lim_{|\bfr|\to r_c}|\psi_B(\bfr)-\psi_0(\bfr)|
/|\psi_0(\bfr)|$.
In such limit, we have $\psi_0(\bfr)
\sim\frac{\Delta r}{r_c}+{\cal O}(kr_c)$, where $\Delta r=|\bfr|-r_c\ll r_c$.
Expanding the Green's function, $G(\bfr,\bfr')$, in the small
$\bfr$ regime, we obtain $\psi_B(\bfr)-\psi_0(\bfr)
= \frac{\Delta r}{r_c}\cdot\frac{\pi^{3/2}}{3\sqrt{5}}
\cdot\frac{M D^2}{\hbar^2 r_c}+\cal{O}({\it kr_c})$. 
As a result, the condition to justify the FBA is 
$\xi=\frac{\pi^{3/2}}{3\sqrt{5}}\frac{a_d}{r_c}\ll 1$ [\oncite{note1}].
For example, we consider the magnetic 
dipolar atom, $^{52}$Cr, with $r_c\sim 100 a_0$
as the typical length scale of van der Waals interaction. 
We find $\xi_{{\rm Cr}}\sim 0.4<1$ and
this explains why results obtained in the FBA
for $^{52}$Cr are comparable to experiments 
[\oncite{dipole_size,dipole_excitation}]. 
However, for polar molecules with 
electric dipole moment of the order of a few Debye, the value of 
$\xi$ can easily be several hundred or more, where a shape resonance
can occur in different channels to breakdown the FBA result.
Therefore, in order to correctly describe the effective
many-body physics of polar molecules, one needs
a self-consistent theory beyond the pseudo-potential, $V_{\rm ps}(\bfr)$,
and the first Born approximation.

\section{Effective Hamiltonian in 3D space}
\label{3D}

\begin{figure}
\includegraphics[width=12cm]{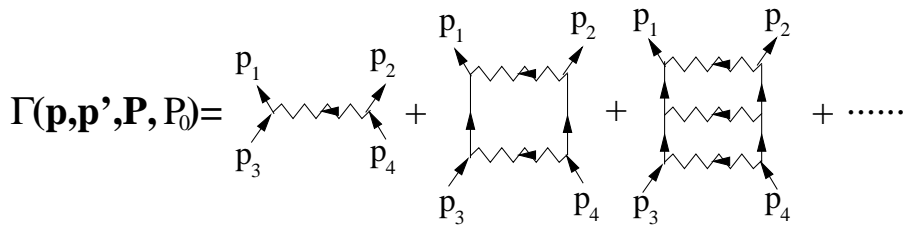}
\caption{
Series expansion for effective interaction in the ladder approximation.
Solid line represents Green's function of bosonic particles and zig-zag 
line is for bare interaction. Here
$\bfp\equiv \frac{1}{2}(\bfp_1-\bfp_2)$ and $\bfp'\equiv
\frac{1}{2}(\bfp_3-\bfp_4)$ are the half the relative momentum, and
$\bfP=\bfp_1+\bfp_2=\bfp_3+\bfp_4$ are the total momentum of the
two scattering particles with frequency $P_0$ (see also 
Ref. [\oncite{Fetter}]).
}
\label{ladder}
\end{figure}
To study the low energy physics of a general dipole interaction
in the many-body medium, one has to use an {\it effective}
two-particle interaction, $\Gamma$, which is just 
the vertex function integrating
out all the contribution of virtual scattering in high energy limit
[\oncite{Fetter}].
A full calculation of the vertex function is usually not available 
(except in some special models of 1D systems), but can be well-approximated
by using the standard ladder approximation 
(see Fig. \ref{ladder}). It is well-known that such ladder approximation
is correct in the low density limit, and is therefore a very suitable 
approximation for systems of dilute cold atoms/molecules.
Following the standard approach to evaluate the Bethe-Salpeter equation
of bosonic particles [\oncite{Fetter}], we can calculate the 
effective two-particle interaction (i.e. vertex function) within
the ladder approximation by using the two-particle scattering amplitude,
$f(\bfp,\bfp')$:
\be
\frac{M}{4\pi \hbar^2}\Gamma(\bfp,\bfp',\bfP,P_0)&=&-f(\bfp,\bfp')
+\frac{4\pi}{\Omega}\sum_{\bfk}f(\bfp,\bfk)f(\bfp',\bfk)^\ast
\left(\frac{1}{\epsilon+2M\mu/\hbar^2-k^2+i 0^+}+
\frac{1}{k^2-p'{}^2-i0^+}\right),
\label{Gamma}
\ee
where $\epsilon=\frac{M}{\hbar^2}\left(\hbar P_0-\hbar^2\bfP^2/4M\right)$ 
is the total kinetic energy in the center-of-mass frame;
$\mu$ is the chemical potential and $\Omega$ is the system volume. 

Using the fact, $f(\bfp',\bfk)^\ast=f(\bfk,\bfp')$, 
the final term of Eq. (\ref{Gamma}) can be evaluated explicitly by
integrating over the solid angle of momentum $\bfk$ in the scattering
amplitude, Eq. (\ref{f_general}). Furthermore, since the partial wave
scattering matrix element, $t_{lm}^{l'm'}(k)$, is known to be 
insensitive to the incident momentum, $k$, in the low energy limit, 
we can also neglect their momentum dependence and replace their value
by a constant, $t_{lm}^{l'm'}(0)$. As a result, the last term of 
Eq. (\ref{Gamma}) can be calculated to be
\wbe
\be
(4\pi)^3\int_0^\infty\frac{k^2 dk}{(2\pi)^3}
\left(\frac{1}{\epsilon+2M\mu/\hbar^2-k^2+i\eta}+
\frac{1}{k^2-p'{}^2-i\eta}\right)
\left[\sum_{l,l',m}\sum_{l''}t_{lm}^{l'' m}(0)
t_{l'' m}^{l' m}(0)Y^\ast_{lm}(\hatp)
Y^{}_{l'm}(\hatp')\right],
\ee
\wee
where we have set $m=m'$ due to the rotational
symmetry about the polarization axis ($z$). 
It is easy to see that the real part of 
the integration cancels out, and the imaginary
part proportional to $\sqrt{2M\mu/\hbar^2}$ in the limit of low energy
scattering ($|p'|,\epsilon\to 0$). 
Therefore, the last term of Eq. (\ref{Gamma}) can be
shown to be negligible when comparing with with the second term, 
$f(\bfp,\bfp')$, in the
low density limit, i.e. ($|t_{lm}^{l'm'}(0)n_{3D}^{1/3}|\ll 1$. 
Here $n_{3D}$ is the 3D particle density).
As a result, in the low energy and dilute limit, one can use
$\Gamma(\bfp,\bfp')=\frac{-4\pi\hbar^2}{M}f(\bfp,\bfp')$ as an effective
``pseudo-potential'' in momentum space (there is no dependence on total
momentum and energy in such limit and we could omit them in $\Gamma$). 
Note that, different from the FBA
used in the literature, we do not have to assume weak bare interaction
in above derivation (strong interaction may still give small
value of scattering matrix element, $|t_{lm}^{l'm'}(0)|$, in the low energy
limit, just as in 
the usual $s$-wave scattering of cold atoms). Complicated electronic
structure and shape resonance effects are all included in the full 
calculation (or experimental measurement) of the matrix elements,
$t_{lm}^{l'm'}(0)$ in all channels. In the rest of this paper, we will study 
the general effective theory and possible new many-body physics 
beyond the FBA {\it without} directly evaluating the scattering 
matrix elements.

Using the derived pseudo-potential (or effective interaction), 
$\Gamma(\bfp,\bfp')=\frac{-4\pi\hbar^2}{M}f(\bfp,\bfp')$, we can write down 
the interacting Hamiltonian in momentum space by using the second 
quantization formalism:
\be
H_I&=&\frac{1}{2\Omega}\sum_{\bfp,\bfp',\bfP} 
\hata^\dagger_{\frac{1}{2}\bfP+\bfp}\hata^\dagger_{\frac{1}{2}\bfP-\bfp}
\hata^{}_{\frac{1}{2}\bfP-\bfp'}\hata^{}_{\frac{1}{2}\bfP+\bfp'}
\Gamma(\bfp,\bfp'),
\label{H_I}
\ee
where $\hata^{}_\bfp$ and $\hata^\dagger_\bfp$ are field operator for bosonic
polar molecules at momentum $\bfp$. The momentum summation from now on is
restricted to low momentum regime as implied by the effective interaction,
$\Gamma(\bfp,\bfp')$. In order to address the effect of pseudo-potential 
beyond the FBA (Eq. (\ref{V_ps})), 
we can divide the contribution of pseudo-potential,
$\Gamma(\bfp,\bfp')$, into three parts: 
\be
\Gamma(\bfp,\bfp')=
\frac{4\pi\hbar^2 a_s}{M}+V_d(\bfp-\bfp')-\frac{4\pi \hbar^2}{M}
f_\Delta(\bfp,\bfp'),
\label{Gamma2}
\ee
where the first term is from the known (dipole moment dependent) 
isotropic $s$-wave scattering, the second term is the usual FBA result for
anisotropic dipolar interaction, the third term, $f_\Delta$, 
is the scattering amplitude deviated from the known FBA results. 
It can be denoted to be
\be
f_\Delta(\bfp_1,\bfp_2)\equiv
-4\pi\sum_{ll'}{}'i^{l'-l}\sum_m \Delta{a}_{ll'}^{(m)}Y_{lm}^\ast(\hat{p}_1)
Y_{l'm}^{}(\hat{p}_2)
\ee
with $\Delta a_{ll'}^{(m)}\equiv 
-i^{l-l'}(t_{lm}^{l'm}(0)-{t}_B{}_{lm}^{l'm}(0))$ being the 
{\it difference} between a full matrix element and its FBA result. 
Here $\sum_{ll'}'$ has excluded $l=l'=0$ term.
Note that in the limit of a weak external field, we have
following orders of magnitudes:
$a_s={\cal O}(1)$, $V_d={\cal O}(D^2)$, and $\Delta a_{ll'}^{(m)}= 
{\cal O}(D^4)$. Therefore, it is easy to see that the pseudo-potential,
$\Gamma$, shown in Eq. (\ref{Gamma2}) has a very smooth connection with
the known FBA results 
[\oncite{dipole_size,dipole_excitation,You_pseudo,Bohn_prl}] 
in the limit of small dipole moment. 
From Eqs. (\ref{H_I}) and (\ref{Gamma2}), it is straightforward to write down
the full effective Hamiltonian to describe the low energy many-body 
physics of polar molecules:
\be
H&=&\sum_\bfp (\epsilon_\bfp-\mu) \hat{a}^\dagger_\bfp \hat{a}^{}_\bfp
+\frac{1}{\Omega}\sum_{\bfp_1,\bfp_2}
\hata^\dagger_{\bfp_1}\hata^{}_{\bfp_2}V_{\rm ext}(\bfp_1-\bfp_2)
\nonumber\\
&&+\frac{1}{2\Omega}\sum_{\bfp_1,\bfp_2,\bfP} 
\hata^\dagger_{\frac{1}{2}\bfP+\bfp_1}\hata^\dagger_{\frac{1}{2}\bfP-\bfp_1}
\hata^{}_{\frac{1}{2}\bfP-\bfp_2}\hata^{}_{\frac{1}{2}\bfP+\bfp_2}
\times\left[\frac{4\pi\hbar^2 a_s}{M}+
V_d(\bfp_1-\bfp_2)-\frac{4\pi\hbar^2 a_s}{M}f_\Delta(\bfp_1,\bfp_2)\right],
\label{H_eff}
\ee
where $V_{\rm ext}(\bfp)$ is the external trapping
potential in momentum space.
Note that Eq. (\ref{H_eff}) has included scattering
from all channels and is also consistent
with the FBA results in the weak dipole limit 
($|f_\Delta|\propto {\cal O}(D^4)$ as $D\to 0$).
When the external electric field is strong enough, there will be 
some modification of the scattering amplitude to be beyond the
results of first Born approximation even in channels different from $s$-wave, 
i.e. $\Delta a_{ll'}^{(m)}\neq 0$ for $l,l'\neq 0$. 
Calculating the magnitude of such modification beyond
the FBA has to be based on the first principle calculation of two scattering
molecules, and
is beyond the scope of this work. Our interest in the current
paper is to study the effective Hamiltonian and the possible many-body
physics when $f_\Delta$ is known. 

In order to compare with the existing theory of weakly interacting dipoles
[\oncite{dipole_size,dipole_excitation,You_pseudo,Bohn_prl}], it
is instructive to express Eq. (\ref{H_eff}) in real space. Details of the 
transformation is shown in Appendix \ref{append}. 
The final result is
\wbe
\be
H&=&
\int d\bfr\,\hatpsi^\dagger(\bfr)
\left[\frac{-\hbar^2\nabla^2}{2m}-\mu+V_{\rm ext}(\bfr)\right]\hatpsi(\bfr)
+\frac{1}{2}\int d\bfr \int d\bfr'
V_{\rm ps}(\bfr-\bfr')\hatpsi^\dagger\left(\bfr\right)
\hatpsi^\dagger\left(\bfr'\right)
\hatpsi\left(\bfr'\right)\hatpsi\left(\bfr\right)
\nonumber\\
&&+
\frac{2\sqrt{\pi}\hbar^2}{M}
\int d\bfR \left[\hatpsi^\dagger(\bfR+\frac{\bfr}{2})
\hatpsi^\dagger(\bfR-\frac{\bfr}{2})\sum_l\Delta{a}_{0l}^{(0)}
\hat{\phi}_{l0}(\bfR)
+{\rm h.c.}\right]
+\frac{2\hbar^2}{M}\sum_{l,l'}{}''\sum_m \Delta {a}^{(m)}_{ll'}
\int d\bfR\,\hat{\phi}_{lm}^\dagger(\bfR) 
\hat{\phi}_{l'm}(\bfR)
\nonumber\\
\label{H_eff_r}
\ee
\wee
where $\sum_{ll'}''$ has excluded $(l,l')=(0,0)$, $(0,2)$ and (2,0),
and we have used the fact that $\Delta a_{ll'}^{(m)}\neq 0$ 
only for $|l-l'|=0,2,4,\cdots$ due to the anisotropic nature 
of dipole potential, $V_d(\bfr)\propto Y_{20}(\hatr)$, and its higher
order effect. As shown in Appendix \ref{append}, we have defined
$\hat{\phi}_{lm}(\bfR)
\equiv\frac{(l+1)!!}{2^{l/2}} \int d\bfr\frac{Y_{lm}(\hat{r})}{r^3}
\hat{\psi}(\bfR+\frac{\bfr}{2})\hat{\psi}(\bfR-\frac{\bfr}{2})$
as a ``pairing'' operator in angular momentum $(l,m)$ channel 
with a spatial ``wavefunction'' $Y_{lm}(\hat{r})/r^3$.
Although such pairing operators do not represent true composite
particles, but can be used to describe the relative motion of two dipoles
before and after scattering: the first term of the last line indicates
an association-dissociation process between a pair and two dipoles, while
the last term describe a transition between ``pairs'' of different
angular momentum channels. These two novel interaction terms should
bring complete new physics in a strongly interacting 
polar molecules, and is worthy for further investigation in the future.

Starting from the effective Hamiltonian, Eq. (\ref{H_eff_r}), we can also
derive the associated Gross-Pitaeviskii type meanfield equation for 
condensate dynamics by using $i\hbar\partial\hatpsi/\partial t=[\hatpsi,H]$
and approximating the bosonic field operator, $\hatpsi(\bfr)$,
to be a $c$-number, $\Psi(\bfr)$. The resulting equation can be written
as following form:
\be
i\hbar\frac{\partial\Psi(\bfr)}{\partial t}&=&\left[\frac{-\hbar^2\nabla^2}{2m}
-\mu+V_{\rm ext}(\bfr)+\int d\bfr' V_{\rm ps}(\bfr-\bfr')|\Psi(\bfr')|^2
\right]\Psi(\bfr)
\nonumber\\
&&+\frac{2\sqrt{\pi}\hbar^2}{M}\int \frac{d\bfr'}{|\bfr'|^3}
\left[\sum_l\frac{(l+1)!!}{2^{l/2}}Y_{l0}(\hatr')\right]
\left[\Psi(\bfr)^\ast\Psi\left(\bfr-\frac{\bfr'}{2}\right)
\Psi\left(\bfr+\frac{\bfr'}{2}\right)
+\Psi(\bfr-\bfr')^\ast \Psi\left(\bfr-\frac{\bfr'}{2}\right)^2
\right]
\nonumber
\\
&&+\frac{2(4\pi)^2\hbar^2}{M}
\int d\bfr_1 \int d\bfr_2\Psi^\ast(\bfr-\bfr_1)
\Psi^{}(\bfr-\frac{\bfr_1+\bfr_2}{2})\Psi^{}(\bfr-\frac{\bfr_1-\bfr_2}{2})
\nonumber\\
&&\times\frac{1}{2}\sum_{ll'm}{}''\Delta{a}_{ll'}^{(m)}
\left[\frac{4\pi}{(2\pi)^3}Y_{lm}^\ast(\hatr_1)
\frac{(l+1)!!}{2^{l/2}}\frac{\pi}{r_1^3}\right]
\left[\frac{4\pi}{(2\pi)^3}Y_{l'm}^{}(\hatr_2)
\frac{(l'+1)!!}{2^{l'/2}}\frac{\pi}{r_2^3}\right]
\label{GP}
\ee
where $\Psi(\bfr)=\langle\hatpsi(\bfr)\rangle$ is the condensate wavefunction.
Similarly, one can also derive associated Bogoliubov-de Genne 
equations for the elementary excitations. We note that the effective
Hamiltonian, Eq. (\ref{H_eff}) and Eq. (\ref{H_eff_r}), and meanfield
equation, Eq. (\ref{GP}), contain
all the effects beyond the simple FBA results, and they will reproduce
the known FBA results when taking $\Delta a_{ll'}^{(m)}=0$.

\section{3D condensate profile}
\label{variational}

\subsection{Gaussian variational wavefunction}

To study the aspect ratio and the stability regime of 
the condensate profile, it is convenient to use the variational
approach [\oncite{bec_book}]. Here we use a 
Gaussian type trial wavefunction,
\be
\Psi(\bfr)&=&\frac{\sqrt{N}\exp(-\rho^2/2R_0^2-z^2/2R_z^2)}
{\pi^{3/4}R_0R_z^{1/2}}
\ee
for the condensate wavefunction in harmonic trapping potential:
$V_{\rm ext}(\bfr)=\frac{1}{2}M\omega_0^2(x^2+y^2)
+\frac{1}{2}M\omega_z^2z^2$, where $\omega_{0}$ and $\omega_z$ 
are the associated trapping frequencies.
Here $N$ is the total number of dipoles,
and $R_0$ and $R_z$ are the Gaussian radii of the condensate in the
$x-y$ plane and along the $z$ axis respectively. 
The variational energy can be obtained easily from the effective Hamiltonian
in the momentum space, Eq. (\ref{H_eff}), via replacing $\hata_\bfk$
by $\Psi_\bfk\equiv\langle \hata_\bfk \rangle=
\frac{1}{\sqrt{\Omega}}\int d\bfr\Psi(\bfr)\,e^{-i\bfk\cdot\bfr}$. We 
therefore obtain
\be
\frac{E(R_0,\beta)}{E_0}&=&\frac{1+2\beta^2}{2\beta^2\tilde{R}_0^2}
+\frac{\tilde{R}^2(2\kappa^2+\beta^2)}{2\kappa^2}
+\frac{2 N \tilde{a}_s}{\sqrt{2\pi}\tilde{R}_0^3\beta}
+\frac{16 A_2(\beta)}{3\sqrt{10\pi}}\frac{N \tilde{a}_d}{\tilde{R}_0^3}
\nonumber\\
&&+\frac{32\beta N}{\sqrt{2\pi}\tilde{R}_0^3}
\left[\sum_{l=2}^\infty
\Delta\tilde{a}_{l,l}^{(0)}A_l(\beta)^2-2\sum_{l\neq l'}^\infty
\Delta\tilde{a}_{l,l'}^{(0)}A_l(\beta)A_{l'}(\beta)\right]
\label{E_MF_a}
\ee
where $A_l(\beta)\equiv\frac{\sqrt{2l+1}}{8}
\int_{-1}^1 dx \frac{P_l(x)}{(1+(\beta^2-1)x^2)^{3/2}}$.
$\beta\equiv R_z/R_0$ and $\kappa\equiv\omega_0/\omega_z$ are the 
condensate and trapping aspect ratios.
We have also scaled all the length scales ($a_s$, $a_d$, 
$\Delta a_{ll'}^{(0)}$ and $R_0$) by the horizontal oscillator length, 
$a_{\rm osc,0}\equiv \sqrt{\hbar/M\omega_0}$ (i.e. $\tilde{R}_0=
R_0/a_{\rm osc,0}$ etc.), and used $E_0\equiv N\hbar^2/2m a_{\rm osc,0}^2$
as the energy scale.

The first two terms in the right hand side of Eq. (\ref{E_MF_a}) are
from the kinetic and potential energies respectively, and the third
is from the $s$-wave scattering channel.
The fourth term in from the contribution 
within the first Born approximation and the second line 
is from the effects beyond the FBA.  
Again we find that the whole meanfield energy of
Eq. (\ref{E_MF_a}) will become the same as calculated within the
First Born approximation by taking $\Delta \tilde{a}_{l,l+2}^{(0)}=0$.
Using the fact that $A_0(\beta)=(4\beta)^{-1}$, 
we find that the contribution of the FBA is of the same form as
the term with $\Delta\tilde{a}^{(0)}_{0,2}$ (both of them are
proportional to $A_2(\beta)$). However, such coincidence
is simply due to the special form Gaussian trial wavefunction. 
Using other trial 
wavefunctions can easily give different aspect ratio dependence 
of these two effects. Besides, we also note that 
$A_l(\beta=1)=0$ for $l\neq 0$, showing that 
for a spherically symmetric condensate ($\beta=1$), 
only $s$-wave scattering channels
are relevant: scattering in finite angular momentum channels are cancelled
out due to spherical symmetry of the condensate profile.
When the condensate profile is highly anisotropic due to external
confinement (say in cigar shape, $\beta\gg 1$ or in pancake 
shape, $\beta\ll 1$), the effects beyond the FBA
will become very crucial.

\subsection{Example: near shape resonance}

For the general form of effective Hamiltonian of
Eqs. (\ref{H_eff}) and (\ref{H_eff_r}),
values of $a_{ll'}^{(m)}$ have to be obtained from the 
first principle calculation [\oncite{You_Born1,Bohn_prl}], which is
however beyond the scope of this paper. In fact, due to the highly
nontrivial inter-molecule interaction in short-distance, the 
low energy scattering matrix element, $t_{lm}^{l'm'}(0)$, can be very 
different from the results of Born approximation in strong dipole
regime. Here we consider the simplest case to study the effect beyond the FBA:
we assume the external electric field is still weak but near the 
first shape resonance regime, where the shape resonance occurs in the
$s$-wave channel so that both $t_{00}^{00}(0)$ and $t_{00}^{20}(0)$ 
are strongly deviated from results in the weak interaction limit.
Scattering matrix elements in other channels are less affected because
of the centrifugal potential for $l\neq 0$. This picture 
is also consistent to the
numerical results shown in Ref. [\oncite{You_Born1}],
where their numerical results of $t_{20}^{40}(0)$ is almost unaffected
by the shape resonance in the $s$-wave channel. (But it does not exclude the
possibility to have significant deviation in other channels in the regime
of much stronger dipolar interaction.)

Under such assumption, we may consider $\Delta a_{0,2}^{(0)}\neq 0$, and
$\Delta_{l,l'}^{(0)}=0$ for all $(l,l')\neq (0,2)$, (2,0) or (0,0).
As a result, the variational energy of Eq. (4) becomes 
(using $A_0(\beta)=1/4\beta$): 
\be
\frac{E(R_0,\beta)}{E_0}&=&\frac{1+2\beta^2}{2\beta^2\tilde{R}_0^2}
+\frac{\tilde{R}^2(2\kappa^2+\beta^2)}{2\kappa^2}
+\frac{2N}{\sqrt{2\pi}\tilde{R}_0^3}
\left[\frac{\tilde{a}_s}{\beta}+8\left(\frac{\tilde{a}_d}{3\sqrt{5}}
-\Delta\tilde{a}_{0,2}^{(0)}\right)A_2(\beta)\right].
\ee
From above result, we find
the contribution of the $\Delta a_{0,2}^{(0)}$ term can reduce 
(since $\Delta a_{0,2}^{(0)}>0$ near the first shape resonance, see
Ref. [\oncite{You_Born1}]) 
the effect of anisotropic feature of dipole interaction.
Although this result is derived from the
Gaussian trial wavefunction, such reduction of anisotropy of the
condensate wavefunction should be still qualitatively
correct for the correct condensate profile.

We note that the ground state energy and the pseudo-potential study
has been also discussed in Ref. [\oncite{Bohn_prl}], where they include
the dipole dependence in the $s$-wave scattering length (i.e. $a_s(D)$) 
and use the FBA results (Eq. (\ref{V_ps})) for the dipole interaction 
near the first few shape resonance. In other words, they did 
not consider the effect of strong deviation of $t_{00}^{20}(0)$ 
from the FBA, which is a very significant result as shown in 
Ref. [\oncite{You_Born1}]. Therefore, the large value of $s$-wave
scattering length near the shape resonance in Ref. [\oncite{Bohn_prl}] 
may have smeared out the contribution of $\Delta a_{0,2}^{(0)}$.
If considering an even stronger dipole moment (larger than the value
for the first few shape resonance), where the scattering amplitudes
may deviate from the FBA result in {\it all} channels, one has to 
solve Eqs. (\ref{H_eff})-(\ref{GP}) with finite values of 
$\Delta a_{l,l'}^{(m)}$ for the correct many-body physics of polar molecules.

\section{Effective Hamiltonian and excitations in 2D:}
\label{2D}

In a 2D homogeneous system, we can assume that the wavefunction in
the $z$ axis is of Gaussian type:
$\phi(z)=\frac{1}{\pi^{1/4}R_z^{1/2}}e^{-z^2/2R_z^2}$, where $R_z$ is
the width of such quasi-2D potential layer. After
integrating out the degree of freedom in $z$ direction (i.e. 
along the direction of external electric field)
of the 3D effective Hamiltonian, Eq. (\ref{H_eff}), we obtain 
\be
H_{\rm 2D}&=&\sum_\bfp\left(\frac{\bfp^2}{2M}-\mu\right)
\hatb^\dagger_\bfp \hatb^{}_\bfp
+\frac{1}{2\Omega_\perp}\sum_{\bfp_1,\bfp_2,\bfq}\left[V_s+V_B(\bfq)\right]
\hatb^\dagger_{\bfp_1+\frac{\bfq}{2}}\hatb^\dagger_{\bfp_2-\frac{\bfq}{2}}
\hatb^{}_{\bfp_2+\frac{\bfq}{2}}\hatb^{}_{\bfp_1-\frac{\bfq}{2}}
\nonumber\\
&&+\frac{1}{2\Omega_\perp}\sum_{\bfp_1,\bfp_2,\bfP}
V_{\Delta}(\bfp_1,\bfp_2)
\hatb^\dagger_{\frac{\bfP}{2}+\bfp_1}\hatb^\dagger_{\frac{\bfP}{2}-\bfp_1}
\hatb^{}_{\frac{\bfP}{2}-\bfp_2}\hatb^{}_{\frac{\bfP}{2}+\bfp_2}.
\ee
where we define $\hatb_\bfp$ and $\hatb_\bfp^\dagger$ to be the field
operator in 2D system with $\bfp$ being the $in-plane$ momentum vector
from now on. $\Omega_\perp$ is the 2D area, and 
$V_s\equiv\frac{4\pi\hbar^2 a_s}{\sqrt{2\pi}\,M R_z}$ is the 
contribution of $s$-wave scattering.
$V_{B}(\bfq)\equiv\frac{\hbar^2 a_d}{M R_z}\frac{4\sqrt{2\pi}}{3}
g\left(\frac{|\bfq| R_z}{\sqrt{2}}\right)$, where $g(x)=1-(3\sqrt{\pi}/2)x\,
e^{x^2}{\rm Erfc}(x)$ with ${\rm Erfc}(x)$ being the 
complementary error function. We also have
\be
V_{\Delta}(\bfp_1,\bfp_2)\equiv\frac{8\sqrt{2\pi}\hbar^2}{MR_z}\sum_{l,l'}'
\sum_m \Delta a_{l,l'}^{(m)}i^{l-l'}F_{lm}^\ast(\bfp_1)
F_{l'm}(\bfp_2)
\ee
to account the contribution beyond FBA, where
\be
F_{lm}(\bfp)\equiv R_z\int dp_z\,e^{-p_z^2R_z^2}Y_{lm}(\hat{p})
=\int_{-\infty}^\infty dx\,e^{-x^2}\sqrt{\frac{2l+1}{4\pi}
\frac{(l-|m|)!}{(l+|m|)!)}}P_l\left(\sqrt{\frac{x^2}
{x^2+p^2R_z^2}}\right)\,e^{im\phi_p},
\ee
with $\phi_p\equiv \tan^{-1}(p_y/p_x)$ being the angle in 2D plane.

At zero temperature, the dipolar
atoms/molecules condense at $\bfp=0$, so that the total energy is
$E_{2D}=\frac{N^2}{2\Omega_\perp}\left[V_s+V_B(0)+V_{\Delta}(0,0)\right]$,
with the chemical potential being
$\mu=n_{2D}\left[V_s+V_B(0)+V_{\Delta}(0,0)\right]$,
where $n_{2D}=N/\Omega_\perp$ is the particle density in the 2D plane.
Keeping only the condensate part ($\hat{b}^{}_0=\hat{b}^\dagger_0
=\sqrt{n_{2D}}$)
and the quadratic order of fluctuations ($\bfp\neq 0$), the effective 
Hamiltonian become:
\be
H_{eff}&=&\sum_\bfp\left(\frac{\bfp^2}{2m}-\mu\right)
a^\dagger_\bfp a^{}_\bfp
+\frac{N}{2\Omega_\perp}\sum_\bfp \left[V_s+V_B(\bfp)+V_{2D}(\bfp,0)\right]
\left(a^\dagger_\bfp a^\dagger_{-\bfp}+a^{}_{-\bfp}a^{}_\bfp\right)
\nonumber\\
&&+\frac{N}{2\Omega_\perp}\sum_\bfp \left[4V_s+2V_B(0)+2V_B(\bfp)
+2V_{\Delta}\left(\frac{\bfp}{2},\frac{\bfp}{2}\right)
+2V_{\Delta}\left(\frac{\bfp}{2},-\frac{\bfp}{2}\right)\right]
a^\dagger_\bfp a^{}_\bfp,
\ee
where we have used the fact that $V_{\Delta}(\bfp_1,\bfp_2)=
V_{\Delta}(-\bfp_1,-\bfp_2)=V_{\Delta}(\bfp_2,\bfp_1)$.
Finally, we could use the Bogoliubov transformation to diagonalize
above Hamiltonian and obtain the following phonon excitation spectrum:
\be
\omega^2_\bfp
&=&\left[\frac{\bfp^2}{2M}
+n_{2D}W_{-}(\bfp)\right]
\left[\frac{\bfp^2}{2m}
+2n_{2D}\left(V_s+V_B(\bfp)+W_+(\bfp)\right)\right],
\ee
where $W_\pm(\bfp)\equiv V_{\Delta}
\left(\frac{\bfp}{2},\frac{\bfp}{2}\right)
+V_{\Delta}\left(\frac{\bfp}{2},-\frac{\bfp}{2}\right)
-V_\Delta(0,0)\pm V_\Delta(\bfp,0)$ accounts the effects beyond
the FBA results.

\begin{figure}
\includegraphics[width=6.5cm]{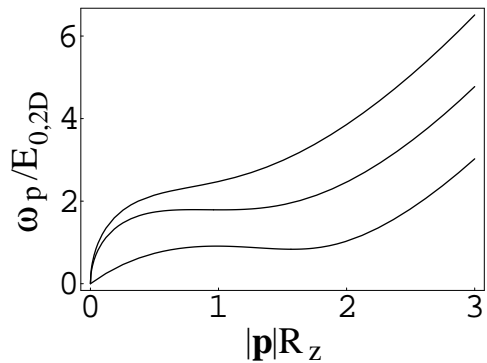}
\caption{
Phonon dispersion, $\omega_\bfp$, as a function of $|\bfp|R_z$
in the 2D system. $n_{2D}\Delta a_{0,2}^{(0)}R_z=0$, 0.05, and 0.1, 
for curves from bottom to top. Other parameters are:
$n_{2D}a_sR_z=0.5$, $n_{2D}a_dR_z=1.4$, and
$E_{0,2D}=\hbar^2/MR_z^2$ is the energy scale.
}
\label{2D_excitation}
\end{figure}
Similar to the 3D case, now we study the situation when only 
$\Delta a_{0,2}^{(0)}\neq 0$ in 
$V_\Delta(\bfp_1,\bfp_2)$, and obtain $V_\Delta(\bfp_1,\bfp_2)=
\frac{-2\sqrt{10\pi}\hbar^2}{MR_z}\Delta a_{0,2}^{(0)}\left[
g(|\bfp_1| R_z)+g(|\bfp_2| R_z)\right]$. The calculated dispersion,
$\omega_\bfp$, for different values of $\Delta a_{0,2}^{(0)}>0$ 
are shown in Fig. \ref{2D_excitation}. 
There are two significant effects to be noted: 
First, in the short wavelength regime,
the roton minimum, predicted [\oncite{dipole_excitation}]
as a feature of dipolar interaction for 2D bosonic polar molecules, 
becomes weaker as
$a_{0,2}^{(0)}$ becomes stronger. Secondly, in the long wavelength limit,
instead of the typical linear dispersion [\oncite{dipole_excitation}], we find 
$\omega_\bfp=C\sqrt{|\bfp|R_z}(1+{\cal O}(|\bfp|R_z))$ with the prefactor
$C=2\sqrt{6}(5\pi^3)^{1/4}\sqrt{\Delta a_{0,2}^{(0)}\left(a_s+\frac{2}{3}a_d
-2\sqrt{5}\Delta a_{0,2}^{(0)}\right)}\times\frac{\hbar^2 n_{2D}}{M R_z}$.
As a result, the
phase fluctuation becomes much stiffer than predicted in the FBA,
showing an enhancement of
the condensate/superfluid density at zero temperature.
More precisely, we can calculate the normal fluid density, $\rho_n$,
according to the transverse current correlation function [\oncite{Nelson}].
The sublinear dispersion of $\omega_\bfp$ gives
$\rho_n(T)=\frac{7!\zeta(7)\hbar^2}{2\pi M}\frac{(k_BT)^7}{C^8R_z^4}$,
which shows a much smaller temperature ($T$) dependence than the result 
obtained for linear dispersion 
($\rho_n=\frac{3\zeta(3)(k_BT)^3}{2\pi M\hbar^2 c_1^4}$
if $\omega_\bfp=c_1\hbar |\bfp|$ [\oncite{Nelson}]). 
According to Landau's two-fluid model and
the universal relation between the 2D superfluid density and 
the Kosterlitz-Thouless transition temperature ($T_c$), 
the superfluid transition temperature ($T_c$) of 2D dipolar system 
is then determined by $k_BT_{c}=\pi\hbar^2\rho_s(T_{c})/2M=\pi\hbar^2(n_{2D}
-\rho_n(T_{c}))/2M$. At temperature below $T_c$, the 
single particle correlation function has a power-law decay with zero
condensate density. These results are also
equivalent to a 2D charged Bose gas ($V(r)=Q^2/r$) [\oncite{2D_C_BEC}]
with an effective charge, $Q=C\sqrt{R_z M/n_{2D}\hbar^2}$.
Such an interesting equivalence implies a possibility to use
neutral polar molecules to simulate a 2D charged boson system
in liquid phase (not doable for ion traps due to the strong
Coulomb interaction and large atom mass), which
may be important to the understanding of the superconducting
Cooper pairs in High $T_c$ thin film [\oncite{high_Tc}].

\section{Summary}
\label{summary}

In summary, we have developed a full effective many-body
theory for 3D and 2D dipolar Bose gases beyond 
the simple first Born approximation. One of the significant consequence is
that the dipolar interaction effect in the 3D condensate can be
reduced near the shape resonance regime. For the 2D system (highly anisotropic
regime), such effect brings a significant change of the 
low energy excitation spectrum.
We believe there should be more interesting results for a polar molecule
system in strong external field regime, where all scattering channels
(besides of the $t_{00}^{20}$ channel) can deviation from the FBA 
significantly. Our 
results therefore are especially important for the future studying of 
the many-body properties of strongly interacting polar molecules.

\section{Acknowledgement}

We thank G. Baskaran, A. Derevianko, D.-H. Lee, 
S. Ronen, L. You, and W.-C. Wu for fruitful discussions. 
Part of this work was done in KITP, Santa Barbara. 
Our work is supported by NSC Taiwan.

\appendix
\section{Effective Hamiltonian in real space}
\label{append}

We note that the single particle 
part, $s$-wave scattering, and the FBA part of interaction
can be easily transformed in to real space as shown in the literature, 
therefore here we just show the results 
for the interaction part beyond the FBA, 
i.e. the last term of Eq. (\ref{H_eff}).
To Fourier transform the effective Hamiltonian, we use 
$\hata_\bfp=\int\frac{d\bfr}{\sqrt{\Omega}}
\hatpsi(\bfr)\,e^{-i\bfp\cdot\bfr}$ and
the interaction part beyond FBA (denoted by $H_I'$) becomes:
\wbe
\be
H_I'&=&\frac{-4\pi\hbar^2}{M}\frac{1}{2\Omega}\sum_{\bfp_1,\bfp_2,\bfP} 
\hata^\ast_{\frac{1}{2}\bfP+\bfp_1}\hata^\ast_{\frac{1}{2}\bfP-\bfp_1}
\hata^{}_{\frac{1}{2}\bfP-\bfp_2}\hata^{}_{\frac{1}{2}\bfP+\bfp_2}
f_\Delta(\bfp_1,\bfp_2)
\nonumber\\
&=&\frac{(4\pi)^2\hbar^2}{M}\frac{1}{2\Omega^3}\sum_{\bfp_1,\bfp_2,\bfP} 
\sum_{ll'm}{}'\Delta{a}_{ll'}^{(m)}
i^{l'-l}Y_{l'm}^\ast(\hatp_1)Y_{lm}(\hatp_2)
\nonumber\\
&&\times
\int d\bfr_1d\bfr_2d\bfr_3d\bfr_4\,e^{i(\frac{1}{2}\bfP+\bfp_1)\cdot\bfr_1}
e^{i(\frac{1}{2}\bfP-\bfp_1)\cdot\bfr_2}
e^{-i(\frac{1}{2}\bfP-\bfp_2)\cdot\bfr_3}
e^{-i(\frac{1}{2}\bfP+\bfp_2)\cdot\bfr_4}
\hatpsi^\dagger(\bfr_1)\hatpsi^\dagger(\bfr_2)
\hatpsi^{}(\bfr_3)\hatpsi^{}(\bfr_4)
\nonumber\\
&=&\frac{(4\pi)^2\hbar^2}{M}\frac{1}{2\Omega^2}\sum_{\bfp_1,\bfp_2} 
\sum_{ll'm}{}'\Delta{a}_{ll'}^{(m)}
i^{l'-l}Y_{l'm}^\ast(\hatp_1)Y_{lm}(\hatp_2)
\nonumber\\
&&\times
\int d\bfR d\bfr_1d\bfr_2\,e^{i\bfp_1\cdot\bfr_1}e^{-i\bfp_2\cdot\bfr_2}
\hatpsi^\dagger(\bfR+\frac{\bfr_1}{2})\hatpsi^\dagger(\bfR-\frac{\bfr_1}{2})
\hatpsi^{}(\bfR-\frac{\bfr_2}{2})\hatpsi^{}(\bfR+\frac{\bfr_2}{2}),
\label{H_I'}
\ee
\wee
where we have integrated out total momentum, $\bfP$, and center of mass
position before scattering, $\bfR'\equiv (\bfr_3+\bfr_4)/2$. 
Note that the summation of angular
momentum quantum number, $\sum_{ll'm}{}'$, has exclude the pure $s$-wave
scattering channel, $l=l'=m=0$.

Now we consider the expansion of plane wave
in spherical harmonic functions:
\be
e^{i\bfk\cdot\bfr}
&=&4\pi\sum_{lm}i^lj_l(kr)Y_{lm}^\ast(\hatr)Y_{lm}(\hatk)
\ee
or equivalently
\be
\int d\Omega_\bfk e^{i\bfk\cdot\bfr} Y_{lm}^\ast(\hatk)
&=&\left\{\begin{array}{lr}
4\pi i^lj_l(kr)Y_{lm}^\ast(\hatr) & \mbox{for }r\neq 0 \\
\sqrt{4\pi}\delta_{l,0} & \mbox{for }r=0
\end{array}\right.
\label{eq1}
\ee
Eq. (\ref{eq1}) suggested that it is more convenient
to separate terms with zero angular momentum quantum number, 
$\Delta a^{(0)}_{0l}=\Delta a^{(0)}_{l0}$, from others
in Eq. (\ref{H_I'}) before carrying out the momentum integral.
(We note that the higher order correction of dipolar interaction, 
$V_d(\bfr)\propto Y_{20}(\hatr)$, can couple $s$-wave to higher moemntum
channels in the strong dipole momentum limit.)
As a result, Eq. (\ref{H_I'}) can be rewritten to be
\wbe
\be
H_I'&=&H_{I1}'+H_{I2}'
\nonumber\\
H_{I1}'&=&\frac{(4\pi)^2\hbar^2}{M}
\int d\bfR d\bfr_1d\bfr_2
\hatpsi^\dagger(\bfR+\frac{\bfr_1}{2})\hatpsi^\dagger(\bfR-\frac{\bfr_1}{2})
\hatpsi^{}(\bfR-\frac{\bfr_2}{2})\hatpsi^{}(\bfR+\frac{\bfr_2}{2})
\nonumber\\
&&\times\sum_l\frac{(-1)^{l/2}\Delta{a}_{0l}^{(0)}}
{2\Omega^2}\sum_{\bfp_1,\bfp_2} 
e^{i\bfp_1\cdot\bfr_1}e^{-i\bfp_2\cdot\bfr_2}\left(Y_{l0}^\ast(\hatp_1)
Y_{00}(\hatp_2)+Y_{00}^\ast(\hatp_1)Y_{l0}(\hatp_2)\right)
\nonumber\\
&=&\frac{(4\pi)^2\hbar^2}{M}\int d\bfR d\bfr_1
\hatpsi^\dagger(\bfR+\frac{\bfr_1}{2})\hatpsi^\dagger(\bfR-\frac{\bfr_1}{2})
\hatpsi^{}(\bfR)\hatpsi^{}(\bfR)\times
\sum_l\frac{(-1)^{l/2}\Delta{a}_{0l}^{(0)}}
{2\sqrt{4\pi}}\left[\frac{1}{\Omega}\sum_{\bfp_1} 
e^{i\bfp_1\cdot\bfr_1}Y_{l0}^\ast(\hatp_1)\right]
\nonumber\\
&&+\frac{(4\pi)^2\hbar^2}{M}\int d\bfR d\bfr_2
\hatpsi^\dagger(\bfR)\hatpsi^\dagger(\bfR)
\hatpsi^{}(\bfR-\frac{\bfr_2}{2})\hatpsi^{}(\bfR+\frac{\bfr_2}{2})
\times\sum_l\frac{(-1)^{l/2}\Delta{a}_{0l}^{(0)}}{2\sqrt{4\pi}}
\left[\frac{1}{\Omega}\sum_{\bfp_2} 
e^{-i\bfp_2\cdot\bfr_2}Y_{l0}(\hatp_2)\right]
\label{H_I1'}
\\
H_{I2}'&=&\frac{(4\pi)^2\hbar^2}{M}
\int d\bfR d\bfr_1d\bfr_2
\hatpsi^\dagger(\bfR+\frac{\bfr_1}{2})\hatpsi^\dagger(\bfR-\frac{\bfr_1}{2})
\hatpsi^{}(\bfR-\frac{\bfr_2}{2})\hatpsi^{}(\bfR+\frac{\bfr_2}{2})
\nonumber\\
&&\times\frac{1}{2}\sum_{ll'm}{}''\Delta{a}_{ll'}^{(m)}
i^{l'-l}\left[\frac{1}{\Omega}\sum_{\bfp_1}e^{i\bfp_1\cdot\bfr_1}
Y_{l'm}^\ast(\hatp_1)\right]\left[\frac{1}{\Omega}\sum_{\bfp_2}
e^{-i\bfp_2\cdot\bfr_2}Y_{lm}(\hatp_2)\right]
\label{H_I2'}
\ee
\wee
where $\sum_{ll'm}{}''$ in the last line 
is a summation excluding any terms with $l$ or $l'=0$.
To get Eq. (\ref{H_I1'}), we have used the fact that $\Omega^{-1}\sum_\bfp
e^{i\bfp\cdot\bfr}=\delta(\bfr)$ and have integrated out one of the 
relative coordinate. Now we can integrated out the solid angle of 
momentum variables ($\bfp_1$ and $\bfp_2$) by using Eq. (\ref{eq1}) for
$l\neq 0$:
\be
\frac{1}{\Omega}\sum_{\bfp}
e^{i\bfp\cdot\bfr}Y_{lm}^\ast(\hatp)&=&\int_0^\Lambda \frac{p^2 dp}{(2\pi)^3}
\int d\Omega_\bfp e^{i\bfp\cdot\bfr} Y_{lm}^\ast(\hatp)
=4\pi i^lY_{lm}^\ast(\hatr)\int_0^\Lambda  \frac{p^2 dp}{(2\pi)^3}j_l(pr)
\label{eq2},
\ee
where $\Lambda$ is the momentum cut-off in atomic length scale 
($\sim r_c^{-1}$), due to the nature of 
effective Hamiltonian obtained by integrating out the high momentum/energy
contribution within the ladder approximation of Eq. (\ref{Gamma}).
In order to regularise it to get a universal expression, we can introduce 
another high momentum cut-off, $e^{-\alpha p}$, inside the integrand and 
taking $\alpha$ to zero ($\alpha\to 0^+$) in the final results. 
Using the fact that only even angular quantum numbers ($l=2m'$) are 
relevant for the scattering between bosonic polar molecules, and applying
the following identity:
\be
\lim_{\alpha\to 0^+}\int_0^\infty p^2 j_{2m'}(pr)e^{-\alpha p}dp&=&
\frac{(2m'+1)!!}{2^{m'}}\frac{\pi}{r^3}
\label{identity}
\ee
for $m'\neq 0$, we can simplify Eq. (\ref{eq2}) further and rewrite the
effective Hamiltonian ($H_{I1}'$ and $H_I2'$) to be
\wbe
\be
H_{I1}'&=&\frac{(4\pi)^2\hbar^2}{M}\int d\bfR d\bfr\left[
\hatpsi^\dagger(\bfR+\frac{\bfr}{2})\hatpsi^\dagger(\bfR-\frac{\bfr}{2})
\hatpsi^{}(\bfR)\hatpsi^{}(\bfR)\times
\sum_l\frac{\Delta{a}_{0l}^{(0)}}{2\sqrt{4\pi}}
\frac{4\pi}{(2\pi)^3}Y_{l0}^\ast(\hatr)\frac{(l+1)!!}{2^{l/2}}\frac{\pi}{r^3}
+{\rm h.c.}\right]
\nonumber\\
&=&\frac{2\sqrt{\pi}\hbar^2}{M}
\int d\bfR \left[\hatpsi^\dagger(\bfR+\frac{\bfr}{2})
\hatpsi^\dagger(\bfR-\frac{\bfr}{2})\sum_l\Delta{a}_{0l}^{(0)}
\hat{\phi}_{l0}(\bfR)
+{\rm h.c.}\right]
\label{H_I1''}
\\
H_{I2}'&=&\frac{(4\pi)^2\hbar^2}{M}
\int d\bfR d\bfr_1d\bfr_2
\hatpsi^\dagger(\bfR+\frac{\bfr_1}{2})\hatpsi^\dagger(\bfR-\frac{\bfr_1}{2})
\hatpsi^{}(\bfR-\frac{\bfr_2}{2})\hatpsi^{}(\bfR+\frac{\bfr_2}{2})
\nonumber\\
&&\times\frac{1}{2}\sum_{ll'm}{}''\Delta{a}_{ll'}^{(m)}
\left[\frac{4\pi}{(2\pi)^3}Y_{lm}^\ast(\hatr_1)
\frac{(l+1)!!}{2^{l/2}}\frac{\pi}{r_1^3}\right]
\left[\frac{4\pi}{(2\pi)^3}Y_{l'm}^{}(\hatr_2)
\frac{(l'+1)!!}{2^{l'/2}}\frac{\pi}{r_2^3}\right]
\nonumber\\
&=&\frac{2\hbar^2}{M}\sum_{l,l'}{}''\sum_m \Delta {a}^{(m)}_{ll'}
\int d\bfR\,\hat{\phi}_{lm}^\dagger(\bfR)
\hat{\phi}_{l'm}(\bfR),
\ee
\wee
where we have defined an effective pairing operator:
\be
\hat{\phi}_{lm}(\bfR)\equiv\frac{(l+1)!!}{2^{l/2}}\int \frac{d\bfr}{|\bfr|^3}
Y_{lm}(\hatr)\hatpsi(\bfR+\frac{\bfr}{2})\hatpsi(\bfR-\frac{\bfr}{2}),
\label{pairing}
\ee
in angular momentum $(l,m)$ channel to simplify the notation.
Therefore, after adding back the known single particle Hamiltonian 
and the FBA results together,
we can obtain the final 
total effective Hamiltonian in real space as shown in Eq. (\ref{H_eff_r}).


\end{document}